\input{aipcheck}

\documentclass[
    ,final            
  ]
  {aipproc}

\layoutstyle{8x11single}
\usepackage{caption}
\usepackage{subfig}
\usepackage{graphicx}

\begin{document}

\title{Jet Tomography at RHIC}


\author{J.C.~Dunlop}{
  address={Brookhaven National Laboratory, Upton, NY 11973 USA}
}

\begin{abstract}
The status of the use of hard probes in heavy ion collisions at 
RHIC is reviewed.  The discovery of strong jet quenching at RHIC 
is a major success.  However, in order to make full use 
of this new phenomenon for full jet emission tomography
of the properties of the collision zone further development
is needed, both experimentally and theoretically.
\end{abstract}

\maketitle


\section{Introduction}
\label{intro}
Jet quenching in nuclear collisions at high energies is
a well-established experimental fact.  The effects are 
not small: hadron spectra at large $p_T$ are suppressed in 
central Au+Au collisions by a factor of four to five relative
to expectations~\cite{Adcox:2001jp,Adler:2002xw,Adler:2003qi,Adams:2003kv}, 
as are the fragments of jets on the 
away-side azimuthally relative to a trigger hadron~\cite{Adler:2002tq}.  
More than three years ago, the lack of suppression
in d+Au collisions definitively proved that the quenching
was due to interactions in the final-state dense medium formed
in Au+Au collisions rather than from depletion of partons
in the initial 
state~\cite{Back:2003ns,Adler:2003ii,Adams:2003im,Arsene:2003yk}.  
This led to statements that the initial gluon density of the
matter produced in central Au+Au collisions
was more than an order of magnitude greater
than that of normal nuclear matter.
For a more detailed description of the state of understanding
a few years ago, see the RHIC ``whitepapers'' from the four 
experimental collaborations~\cite{Arsene:2004fa,Back:2004je,Adams:2005dq,Adcox:2004mh}.

While it is clear that the suppression seen in central
Au+Au collisions requires that the matter is dense, a
more quantitative statement is lacking.
A fundamental problem that arises in some approaches
is that the medium is too black~\cite{Eskola:2004cr,Dainese:2004te}:
the energy loss of partons is so large that one rapidly enters
into a region of diminishing returns, in which the density of the
medium can increase by large factors while the measurable suppression
of the final state hadrons hardly changes.  
This loss of information
is not generally true in all calculational frameworks, and depends on 
the geometry and expansion of the collision
zone, along with the inherent fluctuations from the distribution
of energy loss of the partons.  A recent detailed study
under various scenarios comes to the conclusion that the
tomographic information about the collision zone obtainable
from single hadron suppression is extremely limited~\cite{Renk:2006qg}.
Therefore the challenge is to come up with experimental probes
that recover sensitivity to the properties of the medium.

\begin{figure}[h]
  \centering
      \includegraphics[width=0.45\textwidth]{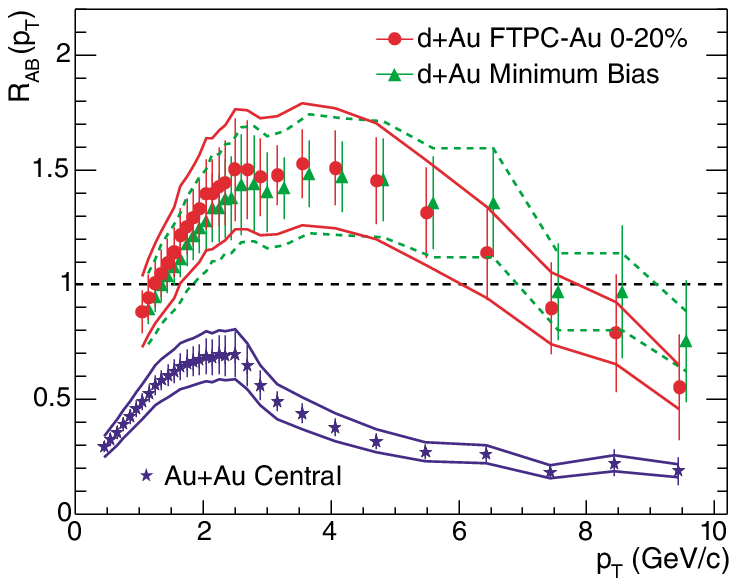}
      \includegraphics[width=0.45\textwidth]{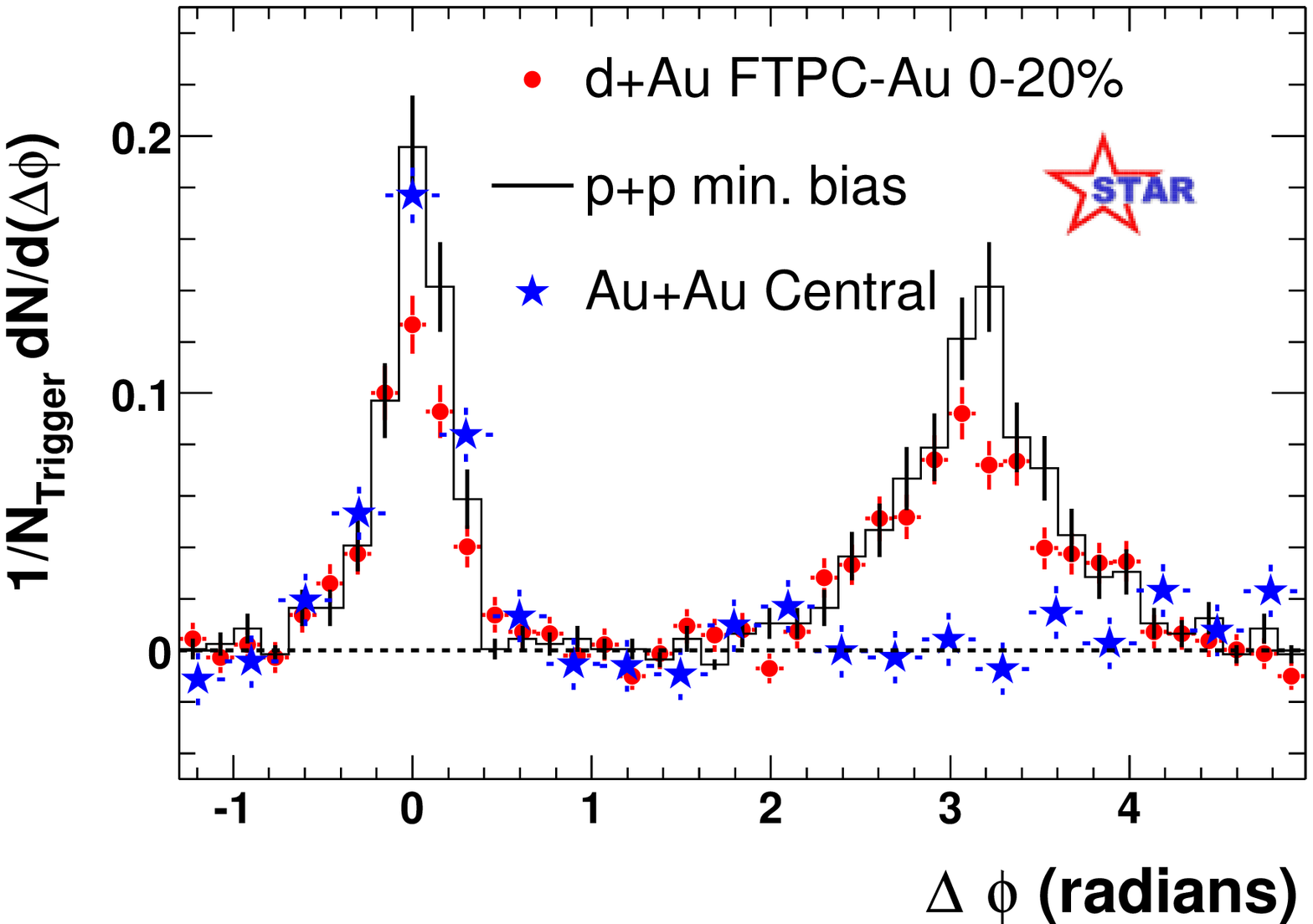}
    \caption{Suppression patterns in Au+Au collisions vs. p+p and d+Au reference
      systems.  Left: Nuclear modification factor $R_{AB}$ for charged 
hadrons in d+Au and central Au+Au collisions.  Figure is from\protect~\cite{Adams:2003im}.  Right:  Azimuthal correlations as conditional yield $1/N_{trig} dN/d\Delta \phi$ for p+p, d+Au, and central Au+Au collisions.  Figure is from~\protect\cite{Adams:2005dq}.} 
    \label{fig:quenching}
\end{figure}

\section{Baseline measurements}
\label{sec:gray}   
The simplest way to determine the properties of a sample is 
to measure the transmission of a probe through
that sample.  This method is used in condensed-matter
physics, and in medical applications such as Positron Emission
Tomography in which the probe is injected directly into the sample.
In order to obtain precise results,
the probe needs to be prepared with 
well-calibrated luminosity and have a 
well-calibrated interaction with the sample.
In the case of jet tomography, the probe is provided
by hard interactions of partons in the initial stages of the
collision, and the calibration of its
luminosity is provided by
measurements in simpler systems, such as p+p and d+Au,
along with theoretical reproduction of these measurements
using perturbative QCD.  Until recently, the calibration
of the interaction of the probe with the medium was taken 
as a given.

\begin{figure}
\centering
\includegraphics[width=0.45\textwidth]{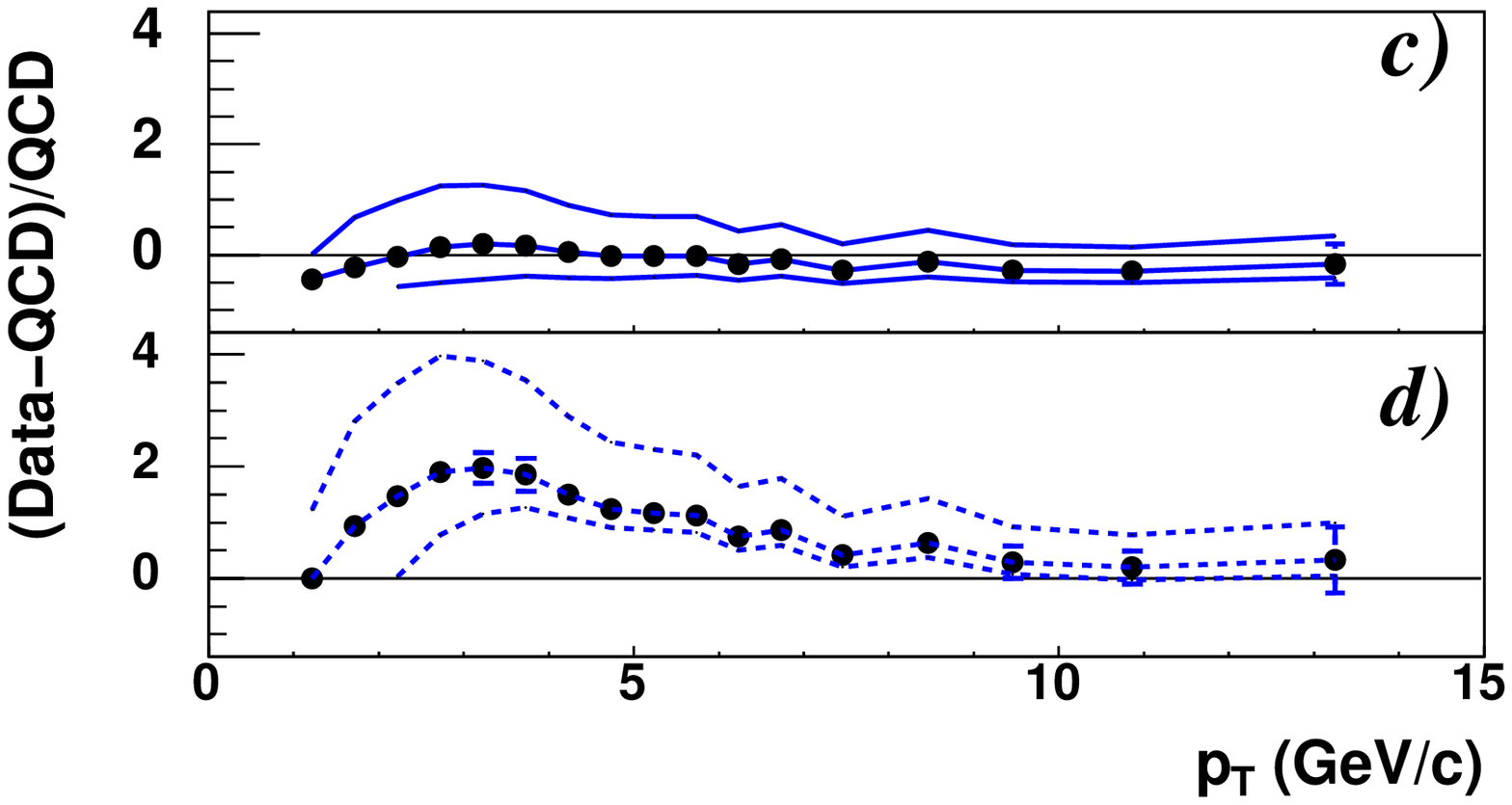}
\includegraphics[width=0.45\textwidth]{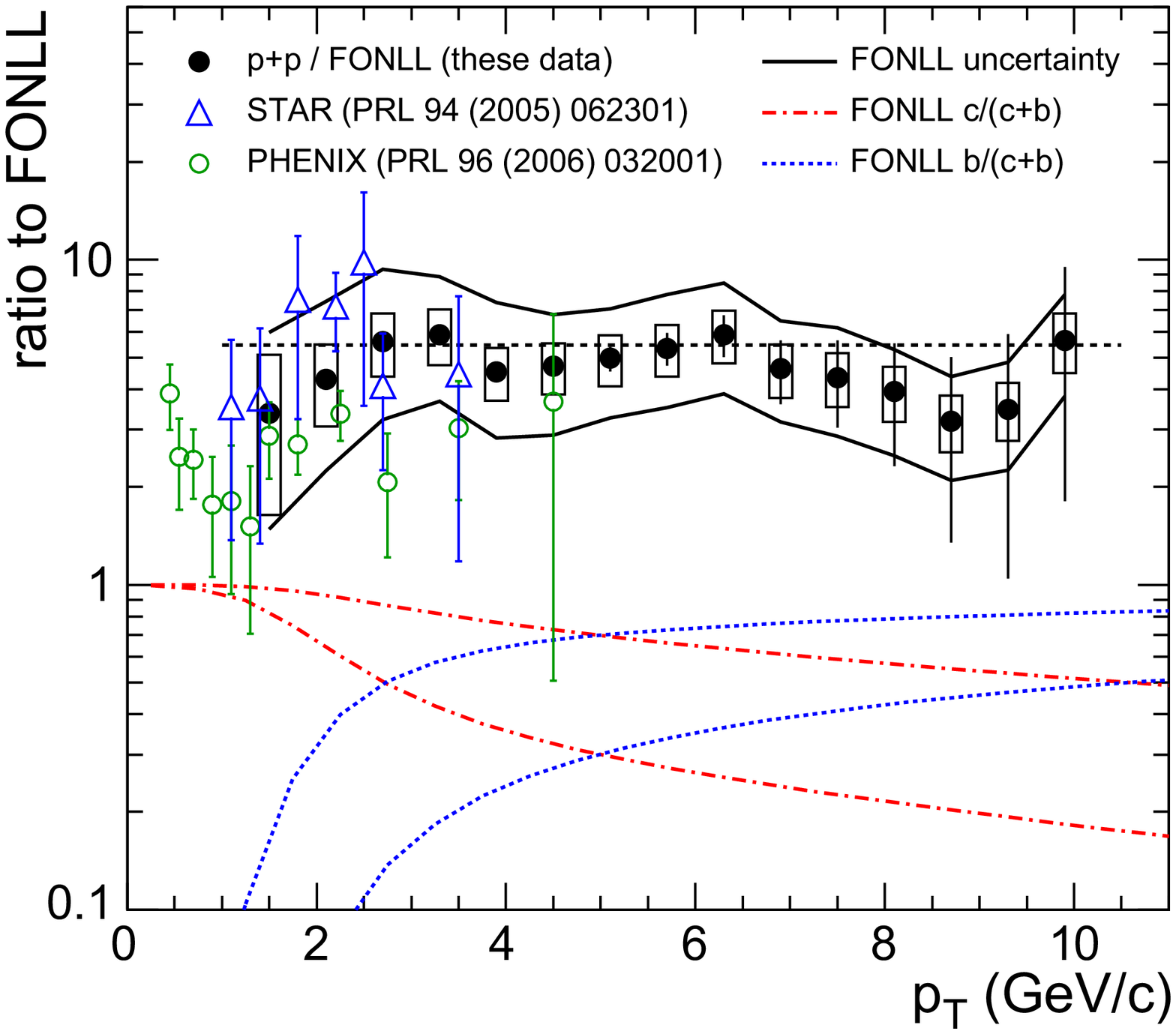}
\caption{Left: Top panel: 
Comparison of $\pi^0$ differential cross-sections in p+p collisions at 
$\sqrt{s}$ = 200 GeV to NLO pQCD calculations.
Top panel incorporates KKP~\protect\cite{Kniehl:2000hk} fragmentation
functions, while bottom panel Kretzer~\protect\cite{Kretzer:2000yf}.
Right: Comparison between non-photonic electrons in p+p collisions
at \protect$\sqrt{s}=$ 200 GeV
and FONLL calculations~\protect\cite{Cacciari:2005rk}. 
 Bands at the bottom indicate allowed
regions of relative contributions to the electrons from charm and bottom.
Figure is from~\protect\cite{Abelev:2006db}, additional data from~\protect\cite{Adler:2005xv,Adams:2004fc}}
\label{fig:nlo_electrons}.
\end{figure}

One can ask to what level the parton-parton luminosity
is calibrated.  One does this by comparing measurements 
with calculations, as shown in figure~\ref{fig:nlo_electrons}.
For light hadrons, the best calculations use the Next to Leading Order (NLO) 
perturbative QCD (pQCD) framework, while for heavy quarks such as 
charm and bottom the best use the Fixed Order Next to Leading Log (FONNL)
framework.  There are three factorized ingredients to these calculations:
the parton distribution functions in the proton, the perturbative 
parton-parton scattering cross-sections, and the fragmentation
functions necessary to convert partons into the observed final-state 
hadrons.  Predictive power is provided by the assumed process-independence
of the ingredients:  parton distribution functions are constrained by measurements
in e+p collisions,
while fragmentation functions are determined from measurements in e+e collisions.
There are significant uncertainties in these calculations
due to knowledge of the incoming parton distribution functions and
the fragmentation functions necessary to convert partons into hadrons.
As an example, 
$\pi^0$ transverse momentum spectra at mid rapidity, shown in figure~\ref{fig:nlo_electrons}, agree 
well with calculations incorporating KKP fragmentation functions~\cite{Kniehl:2000hk}.  Other examples
include $\pi^0$ spectra at forward rapidity~\cite{Adams:2003fx}, proton and
charged pion spectra~\cite{Adams:2006nd}, and direct photon spectra~\cite{Adler:2005qk} at mid-rapidity.
The spectrum of ``non-photonic'' electrons, i.e. those electrons that are not 
from hadrons that decay via processes involving a photons, such as $\pi^0 \rightarrow \gamma \gamma$, 
do not agree with such calculations, as shown in figure~\ref{fig:nlo_electrons}.  
The determination of the relative contribution from charm or bottom decay
to the electrons is additionally
highly uncertain in the FONLL calculations.
These uncertainties can be ameliorated somewhat by accurate measurements
in p+p collisions, along with the observation that the total
integrated yields of charm, measured in the D channel, scale
well with $N_{bin}$\cite{Zhang:2005hi}.  However, the lack of agreement between
theory and experiment leads to complications in interpreting quenching
phenomena in this sector.

\section{Dijets}
\label{sec:correlations}
Correlation measurements, sensitive to dijets,
introduce a different set of 
geometric biases than the suppression of 
single-particle spectra.  Suppression 
in the internal, dense region of the collision
zone biases those hadrons that escape to have
come from hard interactions near the surface
of the collision zone.  Triggering on a hadron,
and then looking at its away-side partner, biases
towards those configurations in which the dijet
emerges tangentially through the system~\cite{Dainese:2005kb}, 
but has the potential to probe deeper into the collision
zone~\cite{Renk:2006nd}.

Recently, STAR has measured true jet-like correlations
on the away-side azimuthally to a trigger hadron~\cite{Adams:2006yt}.
Clear jet-like peaks emerge above background both the side
near and opposite (180 degrees in azimuth) to the trigger hadron.
In previous analyses, the away-side peak was either
so strongly suppressed as to be unobservable over background
~\cite{Adler:2002tq}, or was so strongly widened and softened as to 
make it problematic to call it a collimated ``jet''~\cite{Adams:2005ph}.
This latter low or intermediate $p_T^{associated}$ regime is interesting
in its own right, since in some analyses rather odd structures
are seen~\cite{Adler:2005ee}, and may be a sensitive
way to probe properties of the medium other than its density
~\cite{Casalderrey-Solana:2004qm,Stoecker:2004qu,Ruppert:2005uz}. 
These issues, though, become irrelevant at 
higher $p_T^{trigger}$ and $p_T^{associated}$, and with higher
statistics.

Something provides more information than nothing:
with well-identified peaks, the properties of the peaks
can be studied. 
The conclusion is that, if a dijet is observed,
the fragmentation pattern of the away-side partner to the
jet containing the trigger hadron is unchanged both 
longitudinally along and transverse to the jet axis.
The only modification is that fewer dijets are seen
per trigger hadron.  Interestingly enough, the level
of suppression of the away-side dihadrons is close
to that of the single-particle charged-hadron spectra, 
about a factor of four to five, though these numbers in principle have
little to do with each other.   Such studies have the potential
to recover additional tomographic information,
and are an active area of theoretical investigation.

\section{Gray probes}


If the medium is black to the probe, 
tomographic information is extremely limited, 
which may be true for partons that fragment
into light hadrons such as $\pi^0$.  
The experimental palette is not, however, limited only to such
light hadrons.  By varying the hadron species measured in the
final state, one can vary the parton species used as
a probe, as different species of final-state 
hadrons fragment from different species of partons
that traverse the medium.  Partons of different types are expected
to interact with the medium with different strengths.
Generically, heavy quarks are expected to be less suppressed
than light quarks, which are in turn less suppressed than 
gluons.
Therefore, by varying the parton species one may be able to recover
some of the information lost by the blackness of the medium. 

\begin{figure}
\centering
\includegraphics[width=0.45\textwidth]{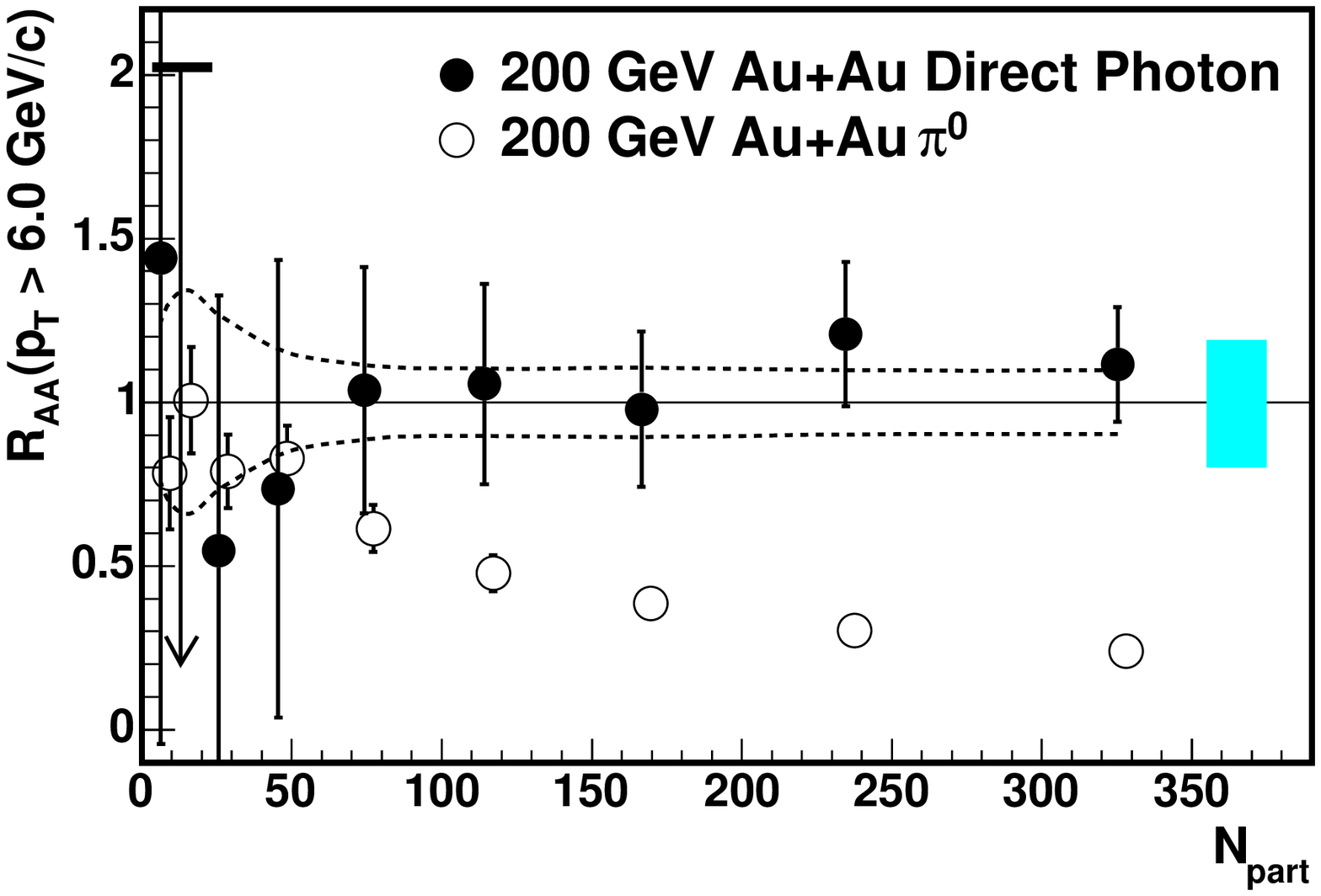}
\includegraphics[width=0.45\textwidth]{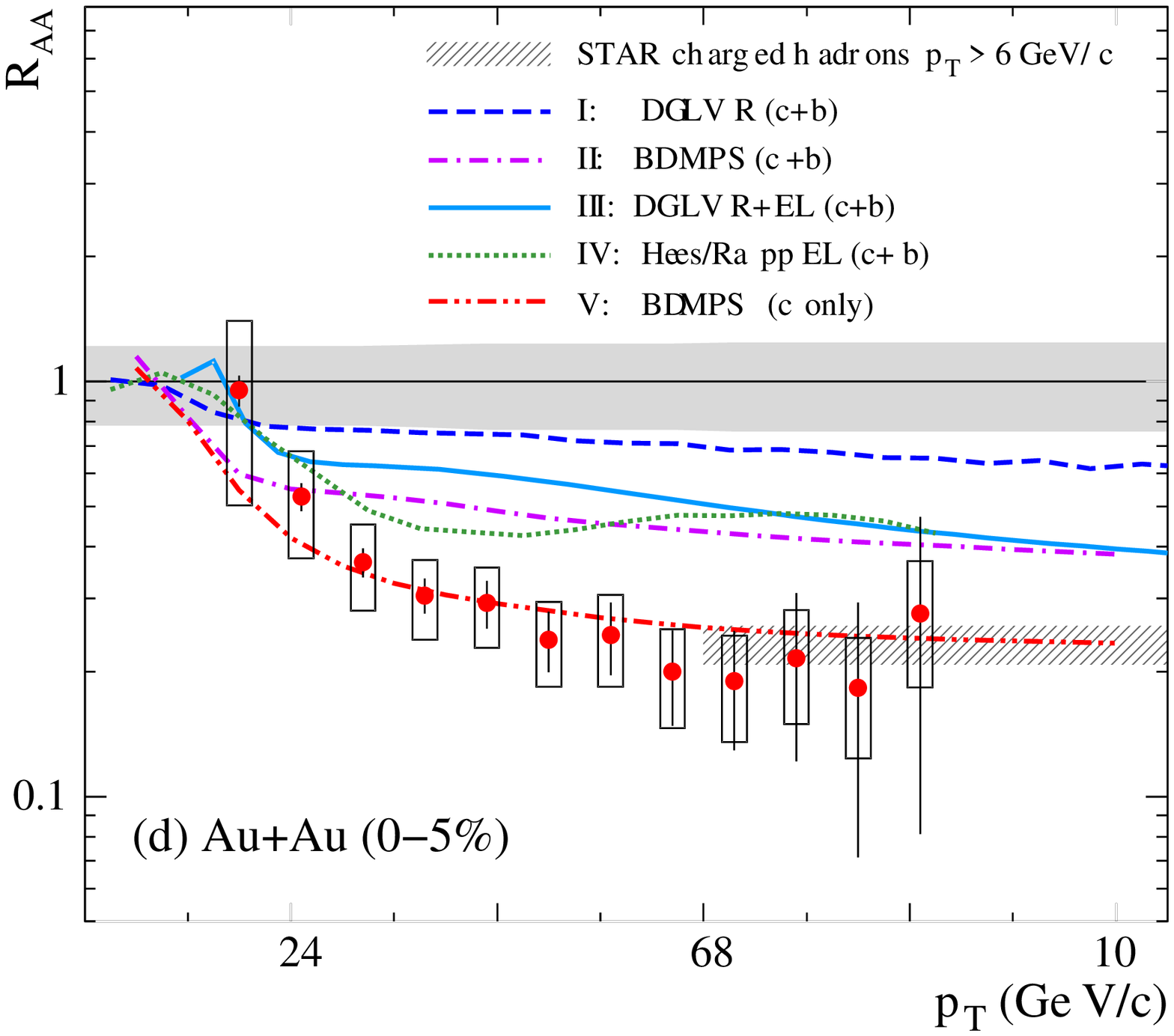}
\caption{Left: Nuclear modification factor $R_{AA}$ for photons
and $\pi^0$ as a function of $N_{part}$ in Au+Au collisions
at $\sqrt{s_{NN}}$ = 200 GeV from PHENIX.  
Figure is from~\protect\cite{Adler:2005ig}.
Right: Nuclear modification factor $R_{AA}$ for 
non-photonic electrons.  Figure is from~\protect\cite{Abelev:2006db}.
}
\label{fig:raa}
\end{figure}

One extreme variation of the interaction of the probe with the
medium is to turn off that interaction entirely, making
the medium transparent to the probe.  A photon produced
in hard parton-parton interactions, through the QCD Compton
diagram, is one such ``white'' probe, as the subsequent
interaction of the photon with the medium is weak.
Direct photons (i.e. 
those photons that do not originate from the decay of hadrons) have 
been measured in Au+Au collisions~\cite{Adler:2005ig}, with 
the result that such photons show no suppression relative
to expectations from next-to-leading-order perturbative QCD 
calculations.  These calculations
also describe results from p+p collisions~\cite{Adler:2005qk}.
This lack of suppression of photons stands in stark contrast
to the large suppression of light hadrons such as $\pi^0$,
as can be clearly seen in figure~\ref{fig:raa}.
The lack of suppression is actually somewhat surprising, as these calculations
indicate that a sizeable fraction of the photons originate
from fragmentation photons, which should in principle be suppressed
as the light hadrons; there are, however, calculations which indicate
that jets passing through the medium can provide an additional
source of photons in central nuclear collisions~\cite{Fries:2002kt}. 
In any case, there is little additional information to be gained
from spectra that are unmodified.

What is really needed is a ``gray'' probe, one that shows some suppression
by the medium but has measurably different suppression than the $\pi^0$.
By changing the final-state hadron measured, and through
this the partonic species used to probe the medium,
one can attempt to find such a probe.

\section{Gluon vs. quark probes}
\label{sec:gluon}
\begin{figure*}
\centering
\includegraphics[width=0.95\textwidth]{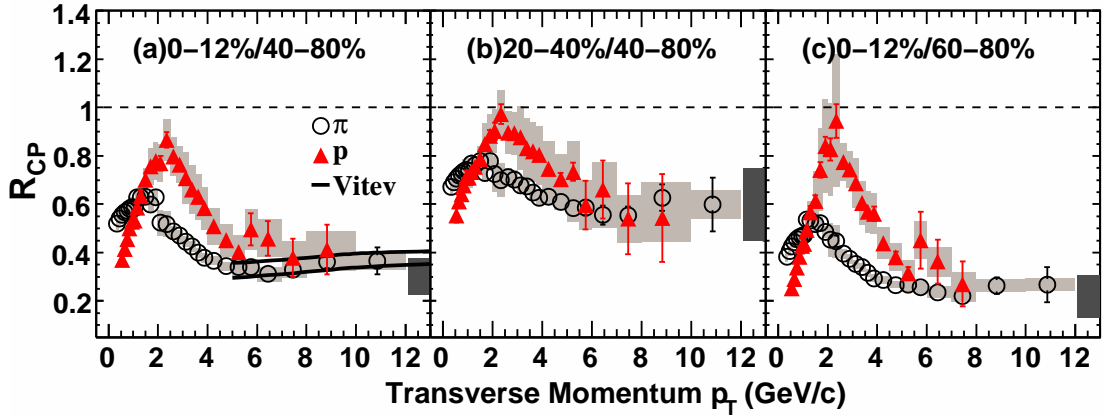}
\caption{Binary-scaled ratio of central to peripheral spectra $R_{CP}$ 
for pions and protons in $\sqrt{s_{NN}}=$ 200 GeV Au+Au collisions
from STAR.  Point-to-point systematic
uncertainties are shown as shaded boxes around the data points.
Dark shaded bands show the normalization systematic uncertainty
in the number of binary collisions $N_{bin}$.
``Vitev'' is a calculation for pions 
including energy loss from gluon bremsstrahlung~\protect\cite{Vitev:2006uc}.
Figure is from~\protect\cite{Adams:2006jr}.}
\label{fig:proton_rcp}
\end{figure*}

Simple Casimir factors in QCD indicate that gluons should interact
more strongly with the medium than quarks; this is borne out by full
calculations~\cite{Wicks:2005gt}.  STAR has recently measured 
charged $\pi$ and proton spectra out to $p_T$ of 10 GeV/c in both 
the simpler d+Au and p+p systems~\cite{Adams:2006nd}, 
and in Au+Au collisions~\cite{Adams:2006jr}.  
Both the pion and proton spectra in the simpler
systems are well described by perturbative QCD calculations,
though in order to describe the proton spectra the AKK fragmentation 
functions~\cite{Albino:2005me}, which separate out the contribution
to final-state hadrons by parton flavor, are necessary.  In this
context, at $p_T$ of 10 GeV/c the contribution of quarks to the production
of pions is significantly larger than that to protons, which remain
produced
dominantly due to the fragmentation of gluons.  Therefore, by measuring
the suppression of protons relative to pions in nuclear collisions, one
is potentially sensitive to differences between light quark and gluon energy 
loss.  The surprising result of the measurement, shown in figure~\ref{fig:proton_rcp}, is that protons and
pions are equally strongly suppressed in central Au+Au collisions.
This is true only for $p_T$ greater than approximately 6 GeV/c, but
the enhancement of baryons in the intermediate $p_T$ region below
this is not at all explainable in a fragmentation framework, 
and indicates interesting physics in its own right.
The equal levels of suppression
may indicate that the medium is equally black to both light quarks
and gluons, and so in the search for gray probes
one needs to find probes that interact less strongly
with the medium than light quarks.

A more discriminating set of measurements will be 
available in the future with the use of photon-tagged
correlations.  Such correlations have long been
seen to have the advantage
that the kinematics of the underlying QCD Compton process
are strongly constrained~\cite{Wang:1996yh}, 
and the recoiling parton is tagged
to be predominantly a quark.
There is the additional advantage that the tagging photon
shines through the collision zone, reducing the geometrical
surface biases that induce the saturation of $R_{AA}$ with increasing
density, perhaps recovering the tomographic information
lost by the blackness of the collision zone~\cite{Renk:2006qg}.  
While first steps have been made towards these 
measurements~\cite{Dietel:2005st}, it is clear that the 
higher luminosities available in the future with the RHIC II 
accelerator upgrade, along with additional experimental work
to subtract backgrounds from fragmentation and decay photons,
are critically needed in order to make definitive measurements
in this channel.
\section{Heavy quarks}
\label{sec:heavy}
Due to their mass, heavy quarks were predicted to interact
less strongly with the medium than light quarks due
to the so-called ``Dead Cone Effect''~\cite{Dokshitzer:2001zm}.
Extended calculations were performed on this effect for both
charm and bottom quarks, for the case of energy loss due to 
gluon radiation~\cite{Djordjevic:2005db,Armesto:2005mz}, with 
the conclusion that the decay products of heavy quarks
should be significantly less suppressed than the fragmentation
products of light quarks.  For charm, the effect could be 
rather subtle, 
though it still remains useful as a clear 
tag for quarks rather than gluons.
The suppression of bottom quarks
is predicted to be significantly smaller than that 
of light quarks in all frameworks. Therefore heavy quarks are a perfect
candidate for a gray probe.

Experimentally, direct reconstruction of charm (D) or bottom (B)
mesons has not been possible in the high $p_T$ regime, though
STAR has directly reconstructed D mesons in both d+Au~\cite{Adams:2004fc}
and Au+Au~\cite{Zhang:2005hi} up to $p_T$ of 3 GeV/c.
However, ``non-photonic'' electrons (i.e. those electrons
that do not arise from decays of lighter mesons such as $\pi^0$
that involve photons or photon conversions)
arise predominantly from decays of B and D mesons, and so
can be used as a proxy.  

Measurements of non-photonic electrons in Au+Au collisions have induced
a crisis.  The medium is not gray to non-photonic electrons:
in fact is it just as black as to light hadrons.  
Figure~\ref{fig:raa} shows
the nuclear suppression factor $R_{AA}$ for non-photonic
electrons as measured by STAR: out to $p_T$ of 8 GeV/c the electrons
are suppressed as strongly as charged hadrons.
This was a major surprise, and has led to significant
questioning of the mechanism of energy loss itself.
The calibration of the interaction of the probe
with the medium, previously taken as a theoretical given,
is currently undergoing major scrutiny. 

The curves in figure~\ref{fig:raa}
show various theoretical attempts to explain the data.
Curve I~\cite{Wicks:2005gt} 
shows a calculation including both charm and bottom
contributions, in which the gluon density is fixed at 
$dN_g/dy$ = 1000 to match
the final-state multiplicity of hadrons.  Curve III~\cite{Wicks:2005gt} 
shows a calculation in the same framework, in which an additional,
collisional, component of energy loss, first
pointed out to be significant
in~\cite{Mustafa:2004dr}, is added.  Curve II~\cite{Armesto:2005iq}
shows a calculation in a different framework, in which the
gluon density is increased to a rather extreme level, but in 
which the dominant source of energy loss remains radiative.
Curve V shows the same calculation, but with the additional
assumption that the bottom contribution to the electron spectra
is negligible.
Curve IV~\cite{vanHees:2005wb} shows a calculation in     
which the energy loss is due to elastic scattering mediated
by resonance excitations (D and B) and LO t-channel gluon
exchange.  Only curve V can reproduce the measurement.
Clearly this measurement provides an extreme challenge to theory.

\section{Conclusion and Outlook}
\label{sec:conclusions}
Where does this leave us in the search for gray probes?
So far the search has failed, as
the nuclear modification factor for all probes accessible
to date is independent of the probe. 
Either there is no gray probe, and there is therefore
little additional tomographic information available
using hadronic probes, or we have not searched
hard enough.  If there is no gray probe, 
it is not at all clear
that a medium so black can be accommodated within
a picture based on perturbative QCD, and so the 
calibration of the interaction of the
probe with the medium would be lost.

There is one possibility remaining: it is still
possible that we have not
measured any beauty in these collisions.
That the FONNL calculation does not reproduce
the measurement in p+p collisions leaves open
the possibility that in the accessible $p_T$ regime
non-photonic electrons
are predominantly from charm.  If this were 
the case, as shown in curve V in figure~\ref{fig:raa},
it would be much easier to accommodate the measured
suppression, and the crisis would be resolved.  

This leads to the future.  Experimentally,
it is critical to measure charm
and bottom separately, both in Au+Au collisions
and in simpler systems.  Ideas have been floated
as to the use of electron-hadron correlations~\cite{Lin:2006rp}
for this purpose, at least in simpler systems.
Both STAR~\cite{Schweda:2005vn} and PHENIX have vertexing
upgrades proposed which will allow
a separation of charm and bottom directly, utilizing
techniques much like those used in high-energy experiments
like CDF and D0.  If the bottom quark is indeed less suppressed
than the other partons, tomographic information will be
recovered, and this, combined with photon-jet correlations,
will allow the technique of jet tomography to enter
into a new, more quantitative stage.

\end{document}